\documentclass[twocolumn,apj]{emulateapj}

\usepackage{color}

\begin{document}

\title{On the origin of the flare emission in \textit{IRIS}' SJI 2832 filter: \\Balmer continuum or spectral lines?}

\author{Lucia Kleint\altaffilmark{1}, Petr Heinzel\altaffilmark{2}, S\"am Krucker\altaffilmark{1,3}}
\altaffiltext{1}{University of Applied Sciences and Arts Northwestern Switzerland, Bahnhofstrasse 6, 5210 Windisch, Switzerland}
\altaffiltext{2}{Astronomical Institute, The Czech Academy of Sciences, Fri\v{c}ova 298, 25165 Ond\v{r}ejov, Czech Republic}
\altaffiltext{3}{Space Sciences Laboratory, University of California, Berkeley, CA, USA}

\begin{abstract}
Continuum (``White-light'', WL) emission dominates the energetics of flares. Filter-based observations, such as the \textit{IRIS} SJI 2832 filter, show WL-like brightenings during flares, but it is unclear whether the emission arises from real continuum emission or enhanced spectral lines, possibly turning into emission. The difficulty in filter-based observations, contrary to spectral observations, is to determine which processes contribute to the observed brightening during flares. Here we determine the contribution of the Balmer continuum and the spectral line emission to \textit{IRIS}' SJI 2832 emission by analyzing the appropriate passband in simultaneous \textit{IRIS} NUV spectra. We find that spectral line emission can contribute up to 100\% to the observed SJI emission, that the relative contributions usually temporally vary, and that the highest SJI enhancements that are observed are most likely because of the Balmer continuum. We conclude that care should be taken when calling SJI 2832 a continuum filter during flares, because the influence of the lines on the emission can be significant.

 \end{abstract}

\section{Introduction}

The optical continuum (``white-light'', WL) emission of flares, which significantly contributes to flare energetics, has been puzzling since its discovery \citep{carrington1859} because of its still unclear origin. The standard flare model predicts that electrons propagate from their low-density coronal acceleration site only until a sufficiently high density is reached, at which point the particles are stopped \citep[``thick target'', e.g.~reviews by][]{benz2008,fletcheretal2011}. It is generally believed that this region lies in the chromosphere. WL emission was found to spatially and temporally coincide with accelerated electrons \citep{kruckeretal2011}, but theory predicts that the WL continuum may partially form in the upper photosphere, which most likely cannot be reached by accelerated electrons. Several explanations have been proposed for this discrepancy, including accelerated protons that may reach lower depths or ``back-warming'' where energy is transported radiatively from chromospheric heights to lower layers \citep[e.g.,][]{metcalfetal1990b}. 

To determine the origin of the continuum emission, it is necessary to either directly measure its height above the limb, which however is quite complex \citep{kruckeretal2015}, or to measure the shape of the spectrum, which allows to deduce the physical mechanism and thus the height of the continuum emission \citep[e.g.,][]{kleintetal2016}. Two mechanisms play an important role for continuum emission in flares: hydrogen recombination and H$^{-}$ emission. Hydrogen recombination in flares arises from a sudden ionization at chromospheric layers, leading to a high abundance of free electrons. It can for example be observed in the Balmer continuum as enhanced emission below 3646 \AA\ during flares. Ground-based observations of the Balmer region reported an increase of the continuum only for some flares \citep[e.g.][]{neidig1989}, and only recently, it became possible to detect the Balmer continuum from space using \textit{IRIS}' NUV wavelength region \citep{heinzelkleint2014}. H$^{-}$ emission is comparably not relevant at NUV wavelengths and at chromospheric heights \citep{kleintetal2016}.

The terminology ``WL flare'' has historically been used for enhancements in the visible continuum, but with the advent of observations in different wavelength ranges, WL is often also used for other (NUV, IR) continua, even though the proper terminology would be e.g.~\ ``NUV continuum''. In this paper, we will use NUV or FUV continuum for \textit{IRIS} observations. An issue of filter-based WL or generally continuum observations is their usually wide spectral throughput, sampling different formation heights, potentially even with contributions from spectral lines that may turn into emission during flares. For example, the ``WL'' filter of the TRACE satellite \citep{metcalfetal2003} spanned from 1700 \AA\ well into the visible range \citep{metcalfetal2003} and therefore covered layers from the photosphere to the transition region without the possibility to assign the observed emission to a certain layer. It is therefore difficult to derive the origin or the mechanism of the WL emission from filter-based observations. \textit{IRIS} contains a ``NUV continuum'' filter with its main transmission around 2828-2832 \AA, but so far it is unclear if this filter is a reliable proxy of the Balmer continuum, or if its emission is influenced by enhanced spectral lines during flares. In this paper, we aim to investigate the contribution of the Balmer continuum and the spectral lines to \textit{IRIS}' 2832 flare emission through the analysis of simultaneously observed spectra.

\begin{figure*}[tb] %
  \centering 
   \includegraphics[width=\textwidth]{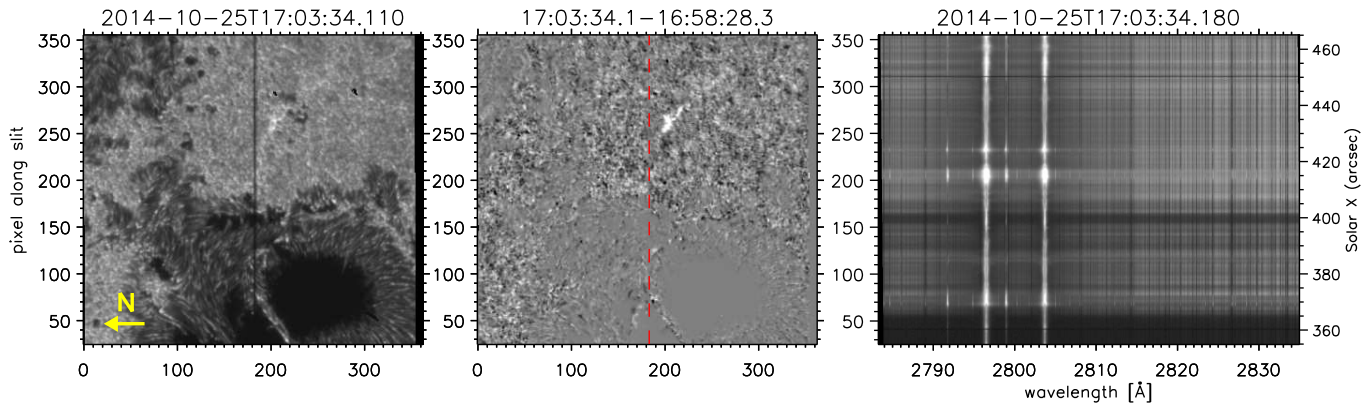}\vspace{-1mm}
   \caption{Left panel: Sample image of the \textit{IRIS} SJI 2832 filter (solar west is up, north is left according to the yellow arrow). Middle panel: difference image of SJI 2832 (times given in title). The red dashed line indicates the position of the slit, bright patches show enhanced continuum and/or line emission during a flare. Right panel: NUV spectrum (scaled to the power of 0.3), showing the full frame readout from 2784 -- 2835 \AA, including the strong \ion{Mg}{2} $k$ and $h$ lines. The lines and continuum are enhanced e.g. around y-pixels $\approx$200--210.}
        \label{overview}
  \end{figure*}
\begin{figure}[tb] %
  \centering 
   \includegraphics[width=.45\textwidth]{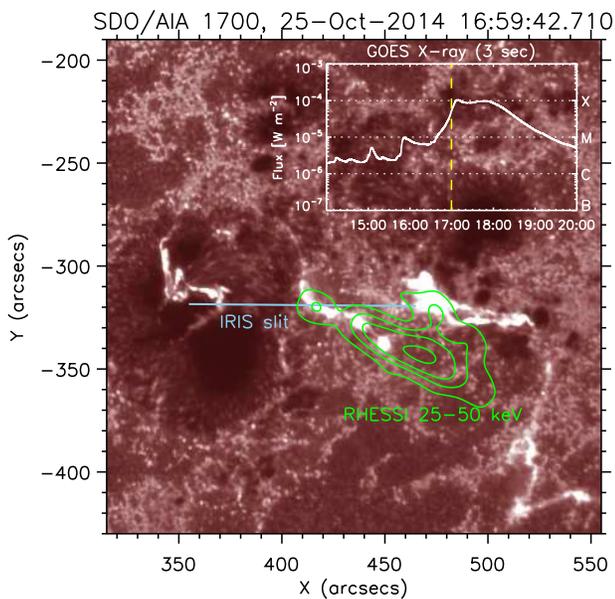}
   \vspace{-.2cm}
   \caption{\textit{IRIS} slit (light blue) drawn on AIA 1700 data during the flare at 16:59:42 UT (yellow dashed line in GOES plot). The inset shows the GOES X-ray 1-8 \AA\ flux with two C-flares before the relatively long-duration X-flare. The green contours show RHESSI emission from 25-50 keV, suffering from pileup effects, but indicating that the \textit{IRIS} slit crossed a footpoint.}
        \label{aiaslit}
  \end{figure}

\begin{figure*}[tb] %
  \centering 
   \includegraphics[width=.77\textwidth]{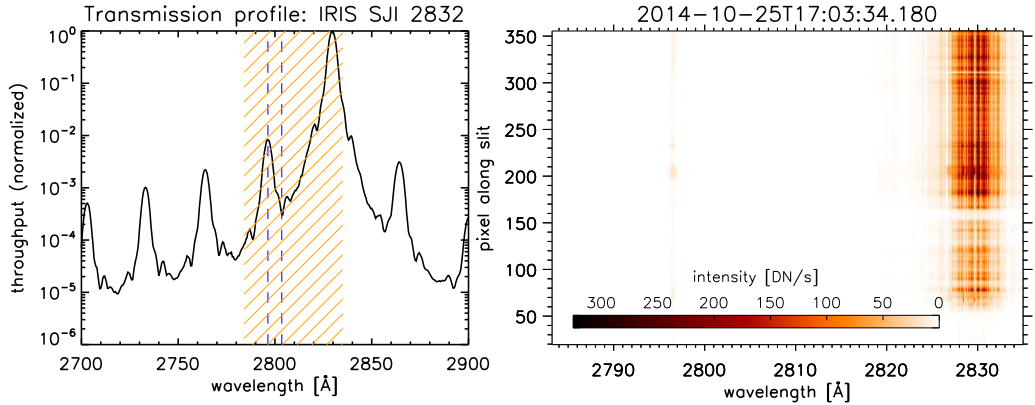}\vspace{-2mm}
   \caption{Left: The normalized effective area of the \textit{IRIS} SJI 2832 filter with its transmission maximum at $\approx$2830 \AA. The vertical dashed lines indicate the wavelengths of the \ion{Mg}{2} $k$ and $h$ lines. The orange shaded area is the NUV spectral range of \textit{IRIS}' spectrograph. Right: \textit{IRIS} NUV spectrum from Fig.~\ref{overview} multiplied with the prefilter transmission to simulate the contribution to the SJI 2832. The color of the spectrum was inverted, but the color scaling is linear according to the colorbar (contrary to Fig.~\ref{overview}).}
        \label{prefilter}
  \end{figure*}

\section{Observations and data reduction}
\label{obs}
The \textit{Interface Region Imaging Spectrograph} \citep[IRIS,][]{iris2014} contains 4 CCDs, two to observe far-UV (FUV) spectra, one for near-UV (NUV) spectra and one for slitjaw images (SJI). Usually, these CCDs are not read out completely, but only selected regions, according to pre-defined linelists. This is advantageous in terms of readout speed and downlink volume. None of the current linelists (small, medium, large, flares) however saves the spectral region around 2828--2832 \AA, which is the region of the main transmission of the SJI 2832 filter. So to compare the emission from SJI 2832 and the NUV spectra, we require an observation with full CCD readout.

To observe an enhanced continuum, it is advantageous to focus on strong flares, which have been observed to show higher continuum emission \citep[e.g.,][]{kuharetal2016}. The NUV Balmer continuum intensity decreases fast and non-linearly during flares, often returning to pre-flare values within less than 150 s \citep{kleintetal2016}. These considerations led to the search for a strong flare with the spectrograph slit passing through the flare ribbon, a full frame readout, and a sit-and-stare observation, giving the highest temporal cadence. We found one \textit{IRIS} flare observation matching the criteria: X1 SOL2014-10-25T17:08 UT.

\textit{IRIS} observed the flare using observation ID 3880106953, a large sit-and-stare, full readout with exposure time 4 s (unchanged throughout the observation), with spatial and spectral binning of 2, resulting in 0\farcs33 pixel$^{-1}$ and 51 m\AA\ pixel$^{-1}$, which ran from 14:58--18:01 UT with the slit oriented east-west. SJI 1330, 2796, and 2832 were recorded, resulting in a 16 s cadence for each filter. 

We used level 2 data, verified the alignment of NUV spectra and SJI using the fiducial lines, and divided the NUV spectra by their effective area to calibrate the wavelength-dependent response of the system. The effective area was normalized to an average value of 1 over the NUV wavelength range before the division, such that the counts do not change significantly.
For our study, we use units of DN/s, but one should be aware that these are now corrected for the wavelength-dependent throughput and not identical the original recorded DN/s, even though the effect is very minor (few percent). We calculated difference images by subtracting spectra/images taken 5 minutes earlier to measure enhancements. An example of the \textit{IRIS} data is shown in Figure~\ref{overview}. The position of the \textit{IRIS} slit is shown on a SDO/AIA 1700 image in Figure~\ref{aiaslit}. Because the NUV data and the SJI data do not overlap perfectly, we plotted the slit from pixels 25-355 (and only analyzed these data). The intersections of the slit with the ribbons are around \textit{IRIS} pixel positions $\sim$70, 208, 354 (left to right on the blue slit), corresponding to $\sim$ 370, 415, 465\arcsec. 

\subsection{Complementary RHESSI data}
The RHESSI hard X-ray observations \citep{linetal2002} reveal a surprisingly weak non-thermal component with very soft spectra and a lack of detected flare photons above ~60 keV. This is atypical for a GOES X-class flare. The peak flux at 35 keV of ~1 photons s$^{-1}$ cm$^{-2}$ keV$^{-1}$ is well below average fluxes \citep{battagliaetal2005} and the soft spectrum with a power law index of around 7 is highly unusual for an X-class flare, which have shown indices of 2-5 in other samples \citep{warmuthmann2016}.
Due to the steep non-thermal spectrum and the small number of high-energy counts, pile-up counts, which are normally hidden in the much stronger non-thermal component, become significant. Pileup occurs when for example two photons of 15 keV reach the detector simultaneously, which are then counted as one photon of 30 keV. Here, pileup counts amount to 40 \% of all counts in the range of 30-50 keV even when the RHESSI attenuators are inserted. Pileup therefore mixes thermal and non-thermal signatures of the flare. This flare therefore is not well suited to study the evolution or position of hard X-ray footpoints to relate them to the Balmer and line emission. Nevertheless, the RHESSI image still shows the location of the flare loop and where emission from the flare ribbon originates. In Figure~\ref{aiaslit} we overplotted RHESSI data as green contours, integrated for 68 s (from 16:59:38-17:00:46 UT) with energy range 25-50 keV and contours of [30,45,60,90]\%, showing the thermal flare loop connecting the flare ribbons, and footpoints along the slit. Because of the steep spectra and the issue with pile up, we cannot unambiguously image the non-thermal component. While the footpoints seen in the 25-50 keV image are likely non-thermal, non-thermal emission could also come from the corona as something seen in so-called thick target coronal sources \citep[e.g.][]{veronigbraun2004}. In summary, the hard X-ray observations of this flare are unusual for an X-class flare and do not show the normally observed strong signature of accelerated electrons indicating that this flare is not an efficient particle accelerator. The influence of this fact on the findings presented in this paper will be discussed in section \ref{disc}.

\section{Comparison of SJI 2832 and NUV enhancements}

Our goal is to compare the enhancement observed along the slit in the SJI 2832 and in the NUV spectra. For the SJI, we calculated the intensity enhancement left and right of the slit (because the slit masks the actual pixel that is observed in the spectra) and averaged the value.

To compare this value to the intensity enhancement observed in the spectra, we need to take into account the transmission profile of the SJI 2832 filter. The SJI 2832 filter has its main transmission (at FWHM) from 2828--2832 \AA, but includes a contribution from the \ion{Mg}{2} $k$ line, the only strong line that may be significant in its passband. The filter's normalized effective area, representing the transmission profile, is shown in the left panel of Fig.~\ref{prefilter}. We multiply this profile with the NUV spectra (example in right panel) and then integrate the resulting emission over the wavelengths to simulate the corresponding SJI 2832 emission.

The time difference ($\Delta t$) of 5 minutes was a compromise between wanting to subtract an image taken before the flare and the fast-changing SJI images, where time differences of more than 5 minutes increased the rms noise. A time difference of 10 minutes was also tested, but did not significantly alter any correlations shown below. The impulsive phase of this flare lasted more than 5 minutes. The enhancements in single pixels in the NUV spectra lasted up to 10 minutes, meaning that during their decay phases spectra from the flare maximum were subtracted when calculating difference spectra for each time step, which gave a spurious apparent negative change of the NUV continuum. We therefore adopted several criteria to only keep trustworthy NUV pixels and time steps for the analysis. We filtered out for each time step
1) all pixels that showed any kind of data issues (values of -200 DN due to lost data, cosmic rays), 2) two pixels that were strongly affected by the small image jitter, making them oscillate between bright penumbral filament and darker penumbra, 3) pixels that showed a negative difference in the \ion{Mg}{2} line emission meaning that our selected $\Delta t$ did not subtract a pre-flare spectrum, 5) pixels with an apparently negative continuum difference (also indicating insufficient $\Delta t$), 6) pixels without a clear step in either continuum or line intensity, indicating slowly evolving solar structures and not flare-related enhancements, 7) pixels with a \ion{Mg}{2} $k$ intensity difference of less than 500 counts, determined empirically to not be related to the flare (during the flare the \ion{Mg}{2} lines reach several thousand counts). We restricted our analysis to 16:31--18:01 UT, fully covering the X1 flare, but omitting the two C-flares that occurred previously.

\begin{figure}[tb] %
  \centering 
   \includegraphics[width=.48\textwidth]{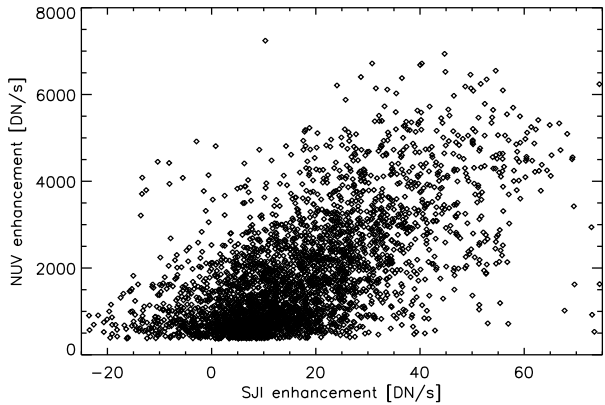}\vspace{-3mm}
   \caption{Scatterplot showing the correlation of the SJI enhancement  (from difference images) versus integrated NUV enhancement (from difference spectra multiplied with the SJI 2832 filter transmission).}
        \label{sjinuv}
  \end{figure}
  
We plot the correlation of SJI and integrated NUV enhancements in Figure \ref{sjinuv}. While the scatter is significant, there is a correlation with coefficient $r$=0.6. The large scatter could be due to the impossibility to compare exactly the same pixel due to the spectrograph slit. In fact, the correlations of NUV enhancements with only either the pixel left of the slit or the pixel right of the slit are significantly worse ($r$=0.4), indicating that the exact spatial position matters. In the following, we will analyze the integrated NUV enhancements as a representative of the SJI enhancements to investigate the influence of the spectral lines and the Balmer continuum on the total SJI enhancement.

\section{Analysis of Line and Balmer Contribution to Total Enhancement}
 \begin{figure}[tb] %
  \centering 
   \includegraphics[clip,bb=20 24 566 525,width=.5\textwidth]{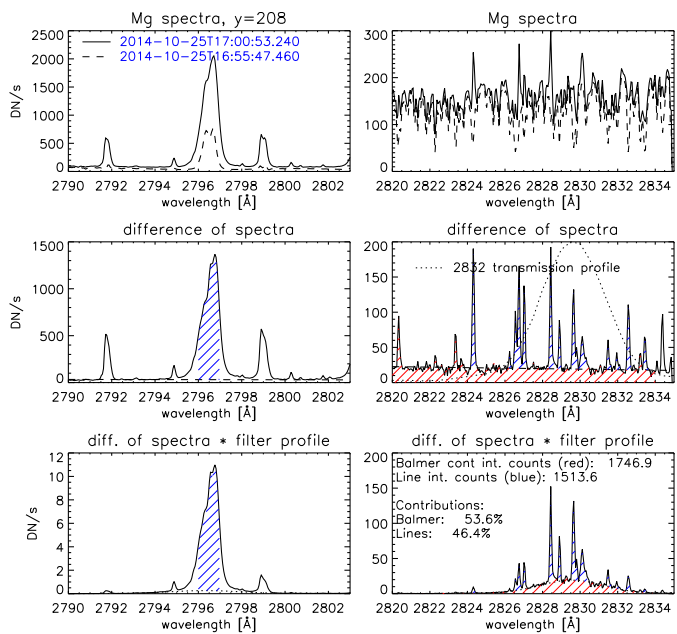}
   \caption{Example of our processing steps to derive the line and Balmer contributions to the continuum: Top row: 2 spectra 5 minutes apart are plotted. Middle row: The difference of these spectra is taken. Bottom row: The difference is multiplied with the SJI 2832 transmission profile. The ratio of the blue areas divided by the red area corresponds to the line emission divided by the Balmer emission.}
        \label{mgdiff}
  \end{figure}
 
To determine whether the spectral lines or the Balmer continuum contribute more to the integrated NUV enhancement, we carry out the steps outlined in Figure~\ref{mgdiff}. The top panel shows two spectra taken 5 minutes apart. The left column shows the \ion{Mg}{2} $k$ line, which because of its large enhancement during flares may have an influence on the SJI 2832 even though its filter transmission is only $\approx$1\% at this wavelength. The right column shows the region in the \ion{Mg}{2} wing around the maximum transmission of the SJI 2832 filter. The middle row shows the difference of the two spectra. Blue shaded areas represent spectral lines (only those significant for the SJI 2832 filter range were marked) and the red shaded area is the overall continuum enhancement, the Balmer continuum. We fitted a linear function (dashed line) through the points outside of spectral lines to determine the continuum level and used this fit in the regions where spectral lines are present to divide the counts into line (above fitted curve) and continuum (below fitted curve) emission. There were cases where the continuum increased, but the line absorption decreased compared to the pre-flare time. In such cases, the continuum counts are derived identically to the case above, but the line counts become a negative number. Since the total enhancement is defined as line plus continuum enhancement, this case may lead to negative percentages (e.g. line decrease of -10\%, continuum increase of 110\%). The third row shows the multiplication with the SJI 2832 filter transmission. We then calculated the total of the Balmer emission (red area) and the total of line emission (blue areas) in DN/s and derived their relative contributions.

 \begin{figure*}[tb] %
  \centering 
   \includegraphics[width=.91\textwidth]{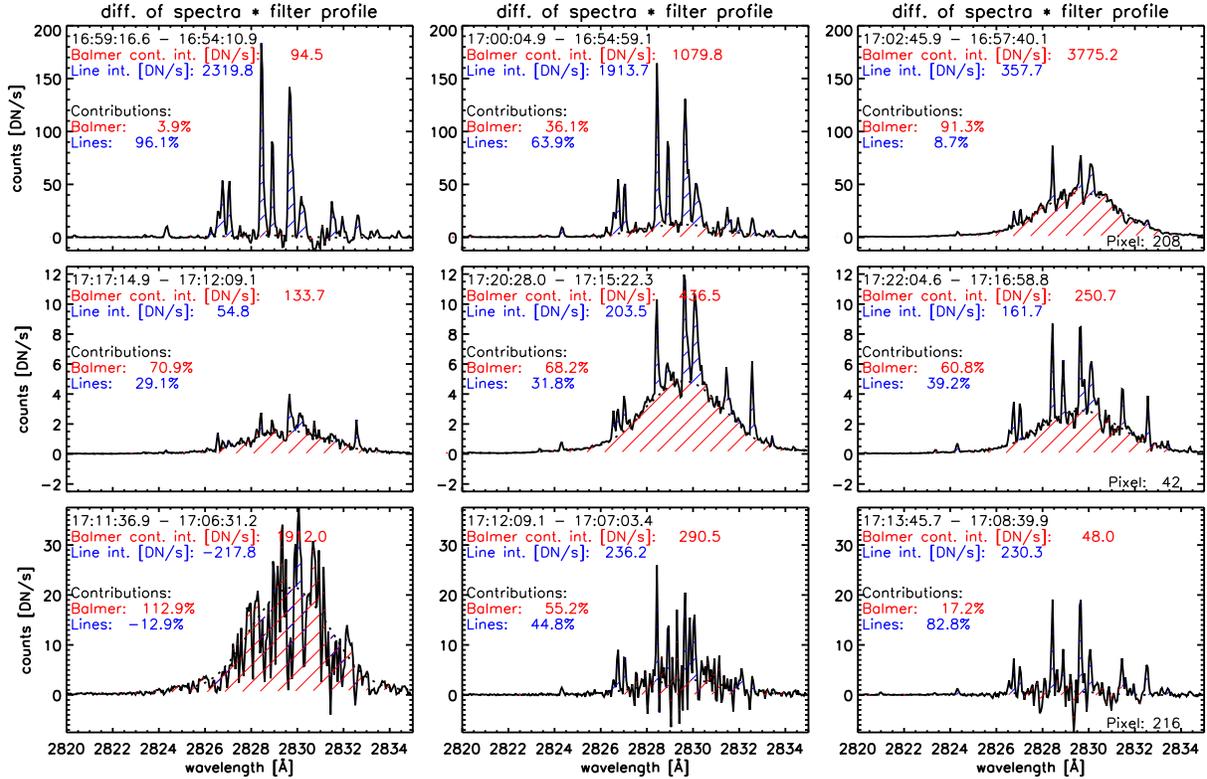}
 \vspace{-.3cm}
   \caption{Temporal evolution of the relative contributions of the line and Balmer emission of three different pixels (rows). Each panel shows a difference of two NUV spectra taken 5 minutes apart (times given in panels) multiplied with the SJI 2832 transmission. Pixel 208 (top row) shows the line emission completely dominating the 2832 filter first, transforming into nearly full Balmer contribution a few minutes later. Pixel 42 showed a constant fraction of the contribution of the two types of emission during an enhancement when the counts temporarily quadrupled. Pixel 216 (bottom) went from Balmer dominated continuum to line emission and later back (not shown). In our sample, such a behavior is more rare than the top and middle panels. 
   }
        \label{pxevol}
  \end{figure*}

  \begin{figure}[tb] %
  \centering 
   \includegraphics[width=.46\textwidth,clip,bb=0 0 425 390]{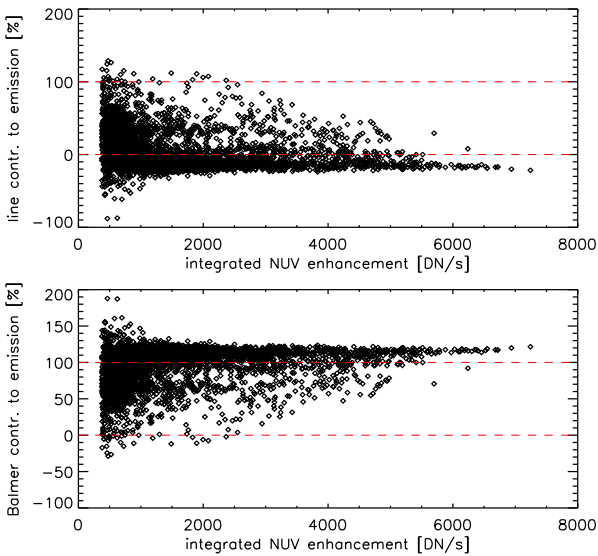}\vspace{-3mm}
   \caption{The relative contributions of the line emission and the Balmer emission to the observed enhanced NUV emission during the flare. Negative line emission means that that the lines' absorption increased during the timeframe.}
        \label{contrfrac}
  \end{figure}

By carrying out this analysis for every pixel and every time step we can study the temporal evolution of the Balmer and line contributions. Figure~\ref{pxevol} shows three time steps (left to right) for three different pixels (top row: pixel 208, middle row: pixel 42, bottom row: pixel 216 along the slit) with different behaviors. The first example shows an evolution from line emission dominating the SJI 2832 passband to 90\% Balmer emission 3.5 minutes later. The second example shows a nearly constant Balmer emission ($\approx$70\%) during 5 minutes. The third example shows a pixel going from Balmer emission to line emission within a minute. One minute after the last panel shown, the Balmer emission increased again in this pixel and dominated the emission. We found all possible behaviors in our sample and there is no correlation of the fraction line/Balmer emission with the flare phase, the \ion{Mg}{2} line emission, or the total enhancement. The only ``rule'' we found is that a very high total NUV enhancement is more likely due to Balmer emission because of the limited number of spectral lines and their limited range of possible increase. This result is shown in Figure~\ref{contrfrac}. The top panel shows the line contribution to the total NUV enhancement (in percent) versus total NUV enhancement. The bottom panel shows the complementary plot, the Balmer contribution. Negative percentages mean that the contribution of the lines or the Balmer continuum decreased during a given time range, for example that the lines showed deeper absorption, while the continuum may have increased at the same time. Because the sum of both plots must be 100\%, some values are above 100\%.

     \begin{figure*}[tb] %
  \centering 
   \includegraphics[width=.96\textwidth]{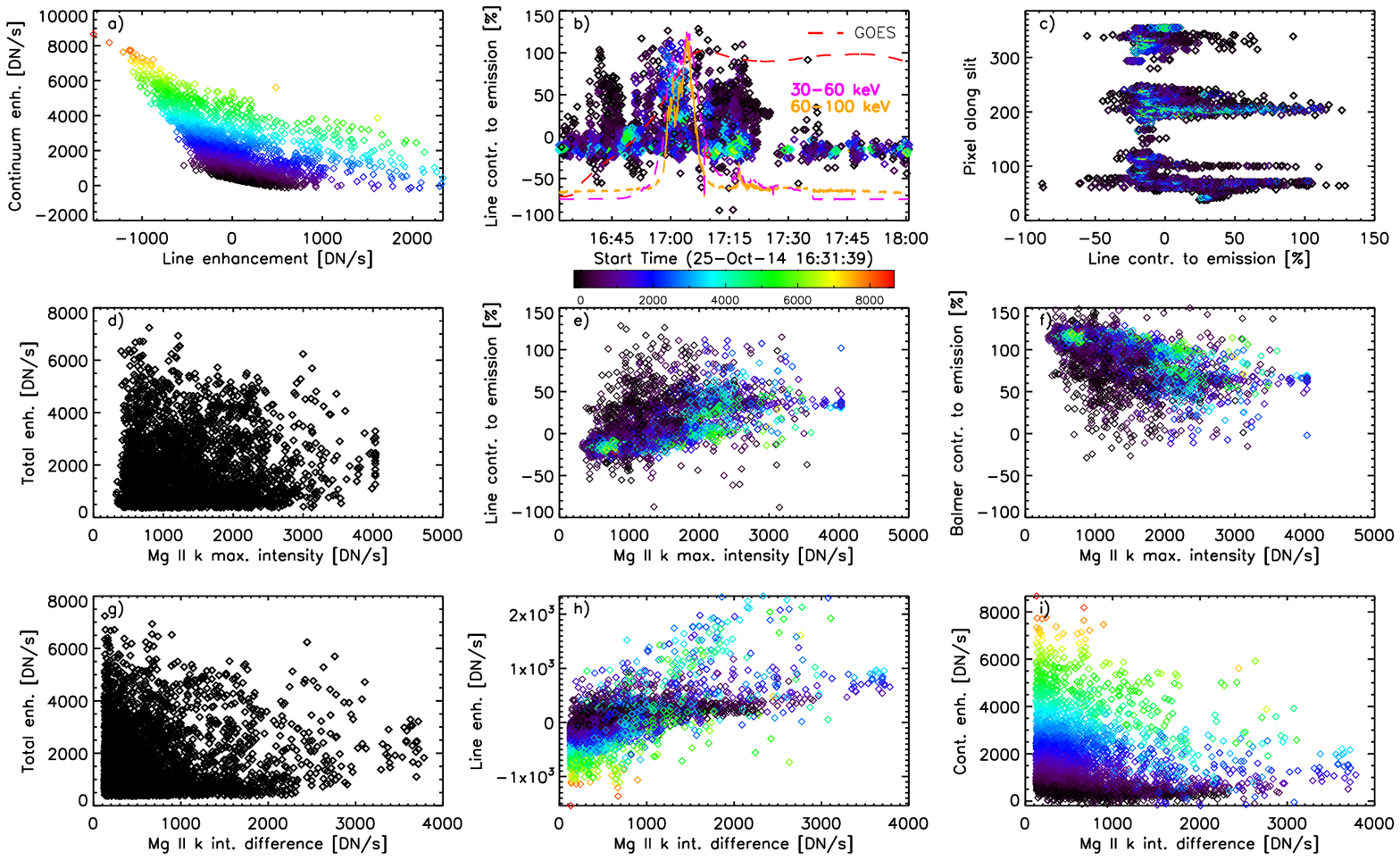}\vspace{-3mm}
   \caption{Scatterplots showing a lack of correlation of the line/Balmer emission to  \ion{Mg}{2} $k$ intensities or intensity differences. The color-coding of the points represents the total intensity enhancement according to the color bar below panel b). See text for more details.}
        \label{panels}
  \end{figure*}
 
     \begin{figure*}[tb] %
  \centering 
   \includegraphics[width=.96\textwidth]{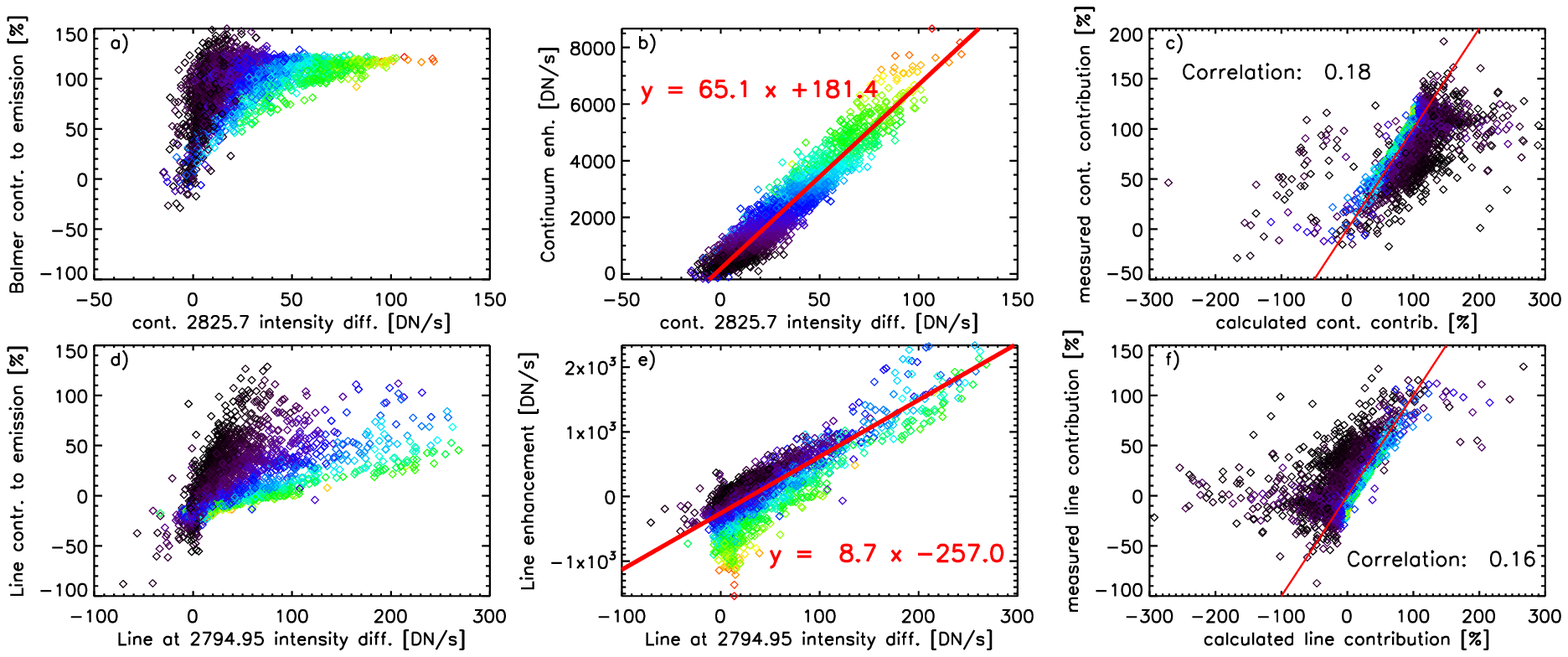}\vspace{-2mm}
   \caption{Correlation of the continuum region averaged between 2825.7 and 2825.8 \AA\ with the integrated continuum enhancement (panel b). The fractional Balmer contribution to the emission is not correlated (panel a). Similarly, a weak line at 2794.95 \AA\ is correlated to the integrated line emission (panel e), but not to the fractional line contribution (panel a). The red lines in b) and e) show linear fits through the correlations.
   By using the linear fits as proxies, it can be investigated if the calculated contributions (x-axes in panels c) and f) ) agree with the contributions from direct measurements of the integrated passbands (y-axes). It is visible that only points with a high integrated NUV enhancement (non-black points in c and f) show a sufficient correlation ($r$=0.7), which means that high enhancements in SJI images may have simple proxies in spectra.}
        \label{panels2}
  \end{figure*}

 \subsection{A proxy for the line/Balmer emission}
 
Now that it is clear that lines can contribute significantly to the observed emission, it is desirable to find a proxy to determine the line/Balmer fraction for observations without the full CCD readout. We therefore attempt correlations of the line and Balmer enhancements with different features that are commonly observed in most linelists.
  Figure~\ref{panels} shows scatterplots of the different quantities. Panel a) shows the line enhancement vs. the Balmer enhancement. The Balmer enhancement can show up to 3 times higher count rates than the line enhancement. Panel b) shows the temporal distribution of the line contribution to the continuum: there are several peaks during the flare and we verified that a single pixel may show multiple peaks. The arbitrarily scaled (but logarithmic) GOES 1-8 \AA\ emission is overplotted in red, showing the maximum of the X1 flare at 17:08 UT. Most of the line contribution occurs in the impulsive phase.
  Panel c) shows that mainly 3 regions along the slit showed significant line emission. The Balmer emission (not plotted) came from the same regions. A comparison with AIA images shows that these 3 regions correspond to intersections of the slit with flare ribbons (see Fig.~\ref{aiaslit}).  The highest line enhancements with $>$1000 DN/s and $>$70\% line contribution to total enhancement all occurred during the times with high HXR emission and all from the pixels around solar X$\approx$415\arcsec. This region corresponds to a HXR footpoint, which may indicate that high line enhancements are correlated to the HXR emission mechanism, but this will need to be verified with a different flare where the IRIS slit crossed more of the HXR emission. Panels d)-f) show the lack of correlation of the line, Balmer or total emission versus the intensity of the  \ion{Mg}{2} $k$ line. This means that a large \ion{Mg}{2} $k$ intensity does not imply a large line contribution to the 2832 SJI. Similarly, the  \ion{Mg}{2} $k$ intensity difference of spectra 5 (or 10) minutes apart is not clearly correlated to 2832 SJI intensity. We additionally tested the correlations by selecting and plotting only a few flare pixels, but the plots remain similar. In summary, \ion{Mg}{2} $k$ does not seem to be a suitable proxy to determine whether a brightening in \textit{IRIS} SJI 2832 is due to Balmer or line emission.

We attempted similar correlations with the subordinate \ion{Mg}{2} lines, with a weak unidentified line at 2794.95 \AA, with an even weaker line at 2800.3 \AA, and with the NUV continuum averaged between 2825.7 and 2825.8 \AA, a continuum region that is commonly observed in the flare line list and was used for continuum studies in \citet{kleintetal2016}. The subordinate \ion{Mg}{2} lines are not correlated to either line or continuum enhancements. 
The NUV continuum region [2825.7, 2825.8] \AA\ shows a direct correlation (correlation factor 0.96) with the continuum enhancement measured from the SJI passband, shown in Figure~\ref{panels2}b. It however is not correlated with the Balmer contribution to emission (panel a). The color coding of the points is identical to Figure~\ref{panels}, according to their total integrated NUV enhancement. The intensity differences of weak lines are correlated to the line emission  in the SJI passband (correlation factor 0.84 for 2794.95 \AA) as shown in panel \ref{panels2}e), though there is no direct correlation to the line contribution to the SJI emission (panel d). By using the correlations in panels b) and e) one can investigate whether one can use them as proxies for the fractional contributions of the line and continuum emission of the SJI emission, which may prove to be useful for example when only the flare line list is used in observations. 

To convert the proxy continuum enhancement at [2825.7,2825.8] \AA\ ($\Delta I_{2826\rm cont}$) and the line intensity enhancement at 2794.95 \AA\ ($\Delta I_{2795\rm line}$) to our measured enhancements from the full-readout analysis, we perform a linear fit (red lines in b and e). Using these two functions we then convert both $\Delta I_{2826\rm cont}$ and $\Delta I_{2795\rm line}$ to values for the enhancement and calculate the line and continuum contributions (x-axis in panels c) and f) ). These contributions can be compared to the measured contributions found from our analysis of the full-frame readout data (y-axis in panels c) and f) ). For a perfect correlation, all points would lie on the red lines in these panels. But the correlation coefficients are below 0.2 for our full sample. It can however be seen that the deviations are mostly caused by black-colored points, i.e. points with a low overall integrated NUV continuum enhancement. If we restrict the analysis to points with a total integrated NUV enhancement above 3500 DN (875 DN/s, i.e. excluding all black points), the correlation coefficient increases to 0.7. This means that the NUV continuum averaged between 2825.7 and 2825.8 \AA\ and the weak line at 2794.95 \AA\ can be used as proxies, provided that the SJI enhancement is bright enough. Due to the relatively large scatter in Fig.~\ref{sjinuv} it is however not possible to provide an absolute number of SJI counts above which the correlation is valid. It also needs to be cautioned that this correlation was tested only for one flare (due to a current lack of other suitable flares) and this flare may not be typical for all flares, at least its X-ray emission was quite unusual.



\section{Discussion and Conclusions}\label{disc}

We have shown that spectral lines have a significant influence on the emission observed by \textit{IRIS}' SJI 2832. It can therefore not be considered a pure Balmer continuum filter during flares. We searched for a proxy to determine when the line contribution dominates or is negligible in SJI 2832 in the absence of NUV spectra of that region. Neither the \ion{Mg}{2} $k$ or the \ion{Mg}{2} subordinate line intensities or intensity enhancements are suitable proxies. Intensity enhancements of weak lines, for example a weak line at 2794.95 \AA, can be used as a proxy for the line emission. The NUV continuum region between 2825.7 and 2825.8 \AA\ can be used as a proxy for the continuum enhancement. This line and continuum region are commonly observed in the flare linelist. It would however need to be verified with future flare observations if the linear relation holds for all flares. If it is proven to be generally valid, then the fractional line and Balmer contributions can be calculated from these proxies, at least for large enough enhancements in the SJI images.

In summary, the Balmer continuum emission can reach higher contributions than the line emission and therefore, very high enhancements observed in SJI 2832 are most likely of Balmer origin. It is however impossible to give a clear cutoff value for the Balmer emission because of the scatter in the SJI-NUV correlation (cf. Figure~\ref{sjinuv}), possibly arising from the fact that neighboring pixels must be used in the SJI for the comparison.

It may be worthwhile to try to correlate the line and Balmer emission to X-ray imaging sources to investigate a possible relation to electron beams. Because this flare's lack of highly energetic electrons and its steep non-themal spectrum, RHESSI imaging was difficult and suffered from pileup of about 40\%, which did not allow us to separate the thermal from the non-thermal emission. Nevertheless, the highest line contributions with strong line enhancements in this flare were co-spatial to a HXR flare footpoint. The low non-thermal counts unfortunately do not allow to study a detailed time evolution. The variability of the line emission indicates variable heating in chromospheric layers on very short time scales, so it would be advantageous to observe a flare with full-frame readout and only SJI 2832 at the highest possible cadence to learn more about flare heating and hopefully have a more typical X-ray spectrum to trace the electron beams.

\acknowledgments
\textit{IRIS} is a NASA small explorer mission developed and operated by LMSAL with mission operations executed at NASA Ames Research center and major contributions to downlink communications funded by ESA and the Norwegian Space Centre. This work was supported by NASA contract NAS 5-98033 for RHESSI. We thank ISSI and ISSI-BJ for enabling interesting discussions on \textit{IRIS} flare observations. PH was supported by the grant No. 16-18495S from the Czech Funding Agency. We thank the referee for constructive suggestions.

\bibliographystyle{apj}
\bibliography{journals,ibisflare}

\end{document}